\documentclass[12pt,preprint]{aastex}

\shorttitle{Cold CO gas in disks}
\shortauthors{Y. Aikawa}

\begin{document}


\title{Cold CO Gas in Protoplanetary Disks}


\author{Yuri Aikawa\altaffilmark{1}}
\affil{Department of Earth and Planetary Sciences, Kobe University,
Kobe 657-8501, Japan}
\email{aikawa@kobe-u.ac.jp}



\begin{abstract}
In a disk around DM Tau, previous observation of $^{13}$CO ($J=2-1$ and $1-0$ transitions) derived the $^{13}$CO gas temperature of $\sim 13-20$K, which is lower than the sublimation temperature of CO (20 K). We argue that the existence of such cold CO can be explained by a vertical mixing of disk material. As the gas is transported from a warm layer to a cold layer, CO is depleted onto dust grains with a timescale of $\sim 10^3$ yr. Because of the steep temperature gradient in the vertical direction, an observable amount of CO is still in the gas phase when the fluid parcel reaches the layer of $\sim 13$ K. Apparent temperature of CO decreases as the maximum grain size increases from $\mu$m-size to mm-size.
\end{abstract}


\section{Introduction}
Theoretical models previously showed that the protoplanetary disks can be divided to three layers in the vertical direction from a chemical point of view \citep{aik99, wl00, aik02, bavb06}. Surface layer of the disk is a PDR (photon-dominated region) because of the interstellar UV, stellar UV, and stellar X-rays. As we go deeper into the disk, the radiation becomes weaker, and the density gets high enough to allow high abundances of organic molecules. In the midplane, on the other hand, heavy-element species are mostly depleted onto grains because of very high density and very low temperature. The three-layer model naturally explains the characteristics of the disk chemistry observed in radio wavelengths: low average abundances of gaseous organic species and relatively high abundances of radical species (ex. CN/HCN ratio) in comparison with the molecular clouds\citep{dut97, qi03, thi04}. Molecular emission lines arise mostly from the intermediate molecular layer, because it is chemically rich and spans a density range that covers critical densities of many (sub-)millimeter rotational transitions.

\citet{dar03} observed two rotational transitions ($J=2-1$ and $1-0$) of various CO isotopes, from which they derived $^{12}$CO temperature to be 30 K and $^{13}$CO temperature $13-20$ K. Since the $^{13}$CO lines trace the deeper layer of the disk than the $^{12}$CO lines, it was the first observational evidence that the gas temperature increases with height from the midplane at radii of $r > 50$ AU, which are traced by the current radio telescopes. In addition, the $^{13}$CO temperature challenges the theoretical model; since the sublimation temperature of CO is $\sim 20$ K, a layer with $T \sim 13-20$ K is considered to be the freeze-out layer.

Recently, \citet{aik06} briefly argued that existence of such cold CO can be explained by vertical mixing due to turbulence. In the vertical direction gas can migrate over $\sim 10$ AU in the vertical direction, and thus can travel from the $T=20$ K layer to the $T=13$ K layer within $10^4$ yr, which is comparable to a freeze-out timescale in the DM Tau disk. In this paper, we examine their arguments in more detail using various disk models. The rest of the paper is organized as follows. In \S 2, we derive a typical migration length of material within the freeze-out timescale of CO. In \S 3, a distance between the $T=13$ K and 20 K layers is measured in various disk models, and is compared with the migration length. We compare our results with the recent numerical calculation of turbulent disk chemistry in \S 4. Summary is contained in \S 5.

\section{Basic Equations}

Following \citet{wil06}, we consider a vertical mixing in the disk. The net transport flux of species $i$ by the turbulent diffusion is
\begin{equation}
\phi(i)=-K n({\rm H}_2)\frac{dx(i)}{dz},
\end {equation}
where $x(i)$ is the abundance of species $i$ relative to the number density of hydrogen molecule $n(i)/n({\rm H}_2)$. We assume that the diffusion coefficient $K$ is equal to the viscosity coefficient: $K\equiv <v_{\rm t}l>=\alpha c_{\rm s} h $ where $v_{\rm t}$ is the turbulent velocity, $l$ is the mixing length, $\alpha$ is the alpha-viscosity parameter \citep{ss73}, $c_{\rm s}$ is the sound velocity, and $h$ is the scale height of the disk. The chemical continuity equation is

\begin{equation}
\frac{\partial n(i)}{\partial t}+\frac{\partial \phi(i)}{\partial z}=
P_{\rm i}-L_{\rm i}
\end{equation}
where $P_{\rm i}$ and $L_{\rm i}$ are the chemical production and loss terms for species $i$. 

We now consider the steady-state abundance distribution of CO near the boundary between the warm molecular layer and freeze-out layer. Since CO is the most abundant and stable species among the carbon-bearing species in the warm molecular layer, the main production and loss terms are desorption from and adsorption onto grains, respectively. As the matter is transported from the warm molecular layer to the freeze-out layer, the desorption term becomes negligible. Hence the vertical distribution of CO is determined by the diffusion from the warm molecular layer and the adsorption onto grains, i.e.
\begin{equation}
\frac{d\phi({\rm CO})}{dz}=-\pi a^2 S \sqrt{\frac{8 k T}{\pi m({\rm CO})}}n_{\rm g} n({\rm CO})
\end{equation}
where $a$ is the grain radius, $S$ is the sticking probability, $k$ is the Boltzman constant, $m$(CO) is the mass of CO molecule, and $n_{\rm g}$ is the grain number density. In order to derive an analytic solution, we neglect the density gradient ($dn({\rm H}_2)/dz$), which is justified because the abundance gradient is steeper than the density gradient. Then the basic equation is reduced to
\begin{equation}
K\frac{d^2 x({\rm CO})}{dz^2}=\pi a^2 S \sqrt{\frac{8kT}{\pi m({\rm CO})}} n_{\rm g} x({\rm CO}).
\end{equation}

The analytic solution for this equation is
\begin{equation}
x=x_0 \exp \left(-\frac{z_0-z}{\lambda}\right)
\end{equation}
where $x_0$ is the canonical CO abundance ($10^{-4}$) in the warm molecular layer and $z_0$ is the height of $T=20$ K layer, i.e. the boundary between the warm molecular layer and freeze-out layer. The decay length $\lambda$ is given by
\begin{equation}
\lambda =\left( \frac{\pi a^2 S}{\alpha c_{\rm s} h}\sqrt{\frac{8kT}{\pi m({\rm CO})}}n_{\rm g}\right)^{-1/2}.
\end{equation}
In other words, material can migrate from the boundary $z=z_0$ to
$z=z_0-\lambda$ within the freeze-out timescale of CO.

\section{Results}
In this section, we evaluate $\lambda$ in various disk models. Considering the vertical temperature gradient, we discuss if the turbulence can transport observable amount of CO to the $T<20$ K layer.

\subsection{Migration length of CO and vertical temperature gradient}
 The decay length $\lambda$ is determined by the ratio between the rate coefficient of adsorption and diffusion coefficient. The former depends on temperature, sticking probability, and total grain surface area. Temperature is set to be 17 K, since we are interested only in a region between 13 K and 20 K, and since the dependence of $\lambda$ on temperature is not strong. Sticking probability $S$ is set to be 0.5. As for dust grains, we assume dust/gas mass ratio of 1\% and grain radius of $10^{-5}$ cm. The number density of grain $n_{\rm g}$ is then $1.6\times10^{-12} n_{\rm H}$, where $n_{\rm H}$ is the number density of hydrogen nuclei. If the grain size distribution follows the MRN model \citep{mrn77}, the total surface area ($\int \pi a^2 n_{\rm g}(a) da$) of dust grains should be about three times larger than assumed in our model. However, observations suggest that the dust grains in disks are larger than the interstellar dust\citep{dal01, rod06}. Dependence of our results on grain size in disks will be discussed in the next subsection. 

The diffusion coefficient depends on $\alpha$ and scale height. We assume $\alpha=0.01$. Analyzing the CO emission lines in DM Tau disk, \citet{gd98} found the intrinsic local velocity dispersion is essentially thermal, with a turbulent component of 0.05 km s$^{-1}$ $< v_{\rm t} < 0.11$ km s$^{-1}$,  while \citet{dar03} estimate $v_{\rm t}\sim 0.16$ km s$^{-1}$. Thus by setting $\alpha=0.01$, we assume the mixing length $l$ of several $10^{-2} h$. The scale height of the disk is given by
\begin{equation}
h=\sqrt{\frac{2 k T r^3}{G M_*\mu m_{\rm H}}},
\end{equation}
where $\mu$ is the mean molecular weight, $m_{\rm H}$ is the hydrogen mass, $G$ is the gravitational constant, and $M_*$ is the mass of the central star. We assume $\mu=2.34$ and $M_*=0.5 M_{\odot}$. Then $\lambda$ is given as a function of hydrogen number density $n_{\rm H}$ and disk radius $r$: 
\begin{equation}
\lambda=\left(1.6 \times 10^{-12}~ \frac{2\pi a^2 S}{\alpha c_{\rm s}}\sqrt\frac{GM_*\mu m_{\rm H}}
{\pi m({\rm CO}) r^3} n_{\rm H} \right)^{-1/2}.
\end{equation}

Here we evaluate $\lambda$ in two disk models from literature: DM Tau disk model, which is constructed to fit the SED assuming the grain size of 0.1 $\mu$m \citep{cec05}, and a model by \citet{dal99}, who assumed interstellar-type dust is well mixed with gas. Models give $n_{\rm H}$ and $T$ as a function of $r$ and $z$. At each radius we calculate $\lambda$ using $n_{\rm H}$ in the layer of $T=13-20$ K. In Figure 1, asterisks, open boxes, and crosses depict $\lambda$ with the density $n_{\rm H}$ at $T=20$ K layer, $T=13$ K layer, and with the average density of $T=13-20$ K layer at each radius, respectively. The black marks are the DM Tau disk model and the gray marks are the \citet{dal99} model. For example, $n_{\rm H}$ is $\sim 10^7$ cm$^{-3}$ at the boundary between the molecular layer and the freeze-out layer at $r\sim 400$ AU in both models. Then the decay length $\lambda$ is $\sim 10$ AU. 

Also shown in Figure 1 (closed circles) is the distance between the layers of $T=20$ K and 13 K, $dz_{\rm 20-13}$, at each radius in the two disk models. It can be seen that at radius of $> 400$ AU, $dz_{\rm 20-13}$ is about 4 times larger than $\lambda$. In other words, when the disk material transported vertically from the 20 K layer to the 13 K layer, gaseous CO is $e^{-4}$ times the canonical value.

It should be noted that the column density of CO gas with $T<20$ K is given by $\lambda n_0({\rm CO})$, where $n_0({\rm CO})$ is the number density of CO molecule at the 20 K layer ($z=z_0$). At the radius of 400 AU, for example, $\lambda$ is about 10 AU, and $n_0$(CO) is $\sim 10^3$ cm$^{-3}$. Then the column density of $^{12}$CO with $T<20$ K is $\sim 10^{17}$ cm$^{-2}$, which corresponds to the opacity of $\tau\sim 1$ for the $J=1-0$ emission line of $^{13}$CO, assuming the isotope abundance ratio of $n(^{12}$CO)/$n(^{13}$CO)$=60$. Therefore, the amount of cold $^{13}$CO produced by the turbulent mixing is high enough to account for the observation by \citet{dar03}.

Finally, it is obvious from the equation (6) that in the disks with larger (smaller) $\alpha$, $\lambda$ is larger (smaller), and thus the cold CO is more (less)abundant.

\subsection{Effect of grain growth}
So far, we have assumed that dust grains in disks are similar to the interstellar grains. However, observations suggest grains in disks are larger than the interstellar grains. Dust lane width of edge-on disks is smaller than that of model disks with interstellar grains. The SED is better reproduced by the disk models with the maximum grain radius in millimeter range \citep{dal01}. In this subsection, we compare the decay length $\lambda$ and the vertical temperature gradient in the disk models of \citet{aik06}. They investigated the effect of grain growth on disk structure ($n_{\rm H}$ and $T$); the number density of grains with radius $a$ is assumed to be proportional to $a^{-3.5}$, and the maximum size $a_{\rm max}$ is varied with the dust/gas mass ratio fixed.

Figure 2 compares $\lambda$ and $dz_{20-13}$ in the models with (a) $a_{\rm max}=10\mu$m and (b) 1 mm. Compared with the case of uniform grain size ($0.1 \mu$m), the total surface area of grains is smaller by a factor of (a) 0.3 and (b) 0.03. Because of the smaller total surface area of grains, the adsorption time scale and thus $\lambda$ are larger in the model with $a_{\rm max}=1$ mm (see eq. 6). On the other hand, $dz_{20-13}$ is smaller in $a_{\rm max}= 1$ mm model than in $a_{\rm max}=10\mu$m model at $r > 100$ AU. Temperature in the disk atmosphere is lower, and hence the vertical width of the disk is smaller in the model with larger $a_{\rm max}$, because of the reduced number of small grains which mainly contribute to the absorption of stellar light. At $r=100$ AU, the temperature in the midplane layer increases with grain growth, and $dz_{\rm 20-13}$ is slightly larger in the $a_{\rm max}=1$ mm model. We can conclude that at $r>100$ AU the amount of cold CO gas increases as the $a_{\rm max}$ increases from $10 \mu$m to 1 mm.

\section{Discussion}
 Recently \citet{sem06} investigated the molecular evolution in a turbulent disk by coupling the chemical reaction network and turbulent diffusion. They showed that the two-dimensional diffusion increases the amount of cold CO, but the amount is too small to account for the observation by \citet{dar03}. Why is the cold CO much less abundant in their model than estimated in the present work?

 We suspect that the deficiency of cold CO is caused by the too efficient grain-surface reactions. Even in the $T\sim 20$ K region in their static model, CO abundance is much smaller than the canonical value. Inclusion of the vertical diffusion {\it reduces} the cold CO, because more CO is converted to H$_2$CO by grain-surface reactions (see their \S 3). Since the sublimation temperature of H$_2$CO is higher than that of CO, the hydrogenation depletes CO at $T\sim 20$ K. The rate of hydrogenation, however, seems to be overestimated in their model, as discussed below.

In the layer of $T\sim 20$ K at $r>100$ AU, we estimate the number density of H atoms (not included in H$_2$) in the gas phase to be $n({\rm H})\lesssim 10$ cm$^{-3}$, referring to the model of \citet{aik06} with $a_{\rm max}=10 \mu$m. The number of H atoms accreting onto a grain with radius $a$ is
\begin{equation}
\sqrt{\frac{8kT}{\pi m({\rm H})}}n({\rm H}) \pi a^2 = 2\times 10^{-4} \left(\frac{T}{20 {\rm K}}\right)^{1/2}\left(\frac{a}{0.1\mu {\rm m}}\right)^2 {\rm s}^{-1}.
\end{equation}
The desorption rate, on the other hand, is
\begin{equation}
\sqrt{\frac{2 n_{\rm s} E_{\rm D}}{\pi^2 m({\rm H})}} \exp\left(-\frac{E_{\rm D}}{kT}\right),
\end{equation}
where $n_{\rm s}$ is the number density of adsorption sites on a grain ($\sim 1.5\times 10^{15}$ cm$^{-2}$), and $E_{\rm D}$ is the adsorption energy \citep{her93}. If $E_{\rm D}$ is 350 K \citep{hh93}, the desorption rate is $7.4\times 10^4$ s$^{-1}$ at 20 K. Hence a number of H atoms on a dust grain is much smaller than unity even for grains with $a=10 \mu$m. With such a low H atom abundance on a grain, the hydrogenation rate is limited by the accretion rate rather than determined by a migration rate on a grain \citep{cas98}. According to the model description of \citet{sem06}, they seems to have estimated the hydrogenation rate using the migration rate, which can be an overestimate.

Reality can be more complex than the above argument. For example,
grains are not spherical
but would be porous, which will enhance the possibility of reaction before
desorption \citep{pb06}. On the other hand, laboratory experiment suggests
that CO is hydrogenated only in the surface layer of the ice mantle
\citep{wsk03}. Most of the currently-available astrochemistry models
may overestimate the hydrogenation rates, since they do not discriminate the
surface species from species inside ice mantle.
Modeling the grain-surface reactions is still a challenging subject in
the astrochemical studies.

However, it is an observational fact that cold ($\sim 13$ K) CO gas exists
in the disk. If the CO depletion is caused
by hydrogenation in the model by \citet{sem06}, it tells us even the
abundance of simple gaseous species can be dependent on how we model the
grain-surface reactions -- an important caution for future models.

\section{Summary}
 We have shown that in protoplanetary disks the turbulent diffusion can produce observable amount of CO that is colder than the sublimation temperature. As disk material is transported to the layer with $T<20$ K, freeze out proceeds with timescale of $\sim 10^3 (n_{\rm H}/10^7 {\rm cm}^{-3})$. Since the gas migrates over a distance of $\sim 10$ AU within this timescale, CO gas of column density $\sim 10^{17}$ cm$^{-2}$ exists in the cold ($T<20$ K) layers, which explains the observation by \citet{dar03}. As the dust grains in the disk become larger by coagulation, the freeze-out time scale increases, and the vertical temperature gradient at $T\sim 20$ K gets steeper. Hence, the apparent CO temperature decreases as the maximum grain size increases from $\mu$m-size to mm-size.


\acknowledgments

I thank referees for their comments which were helpful in improving
the manuscript. I am grateful to C. Ceccarelli, C. Dominik, P. D'Alessio
and H. Nomura for providing their disk models.
This work is supported by a Grant-in-Aid for Scientific Research (16036205,
17039008) and ``The 21st Century COE Program of Origin and Evolution of
Planetary Systems" of the Ministry of Education, Culture, Sports, Science
and Technology of Japan (MEXT).

\clearpage

\begin{figure}
\plotone{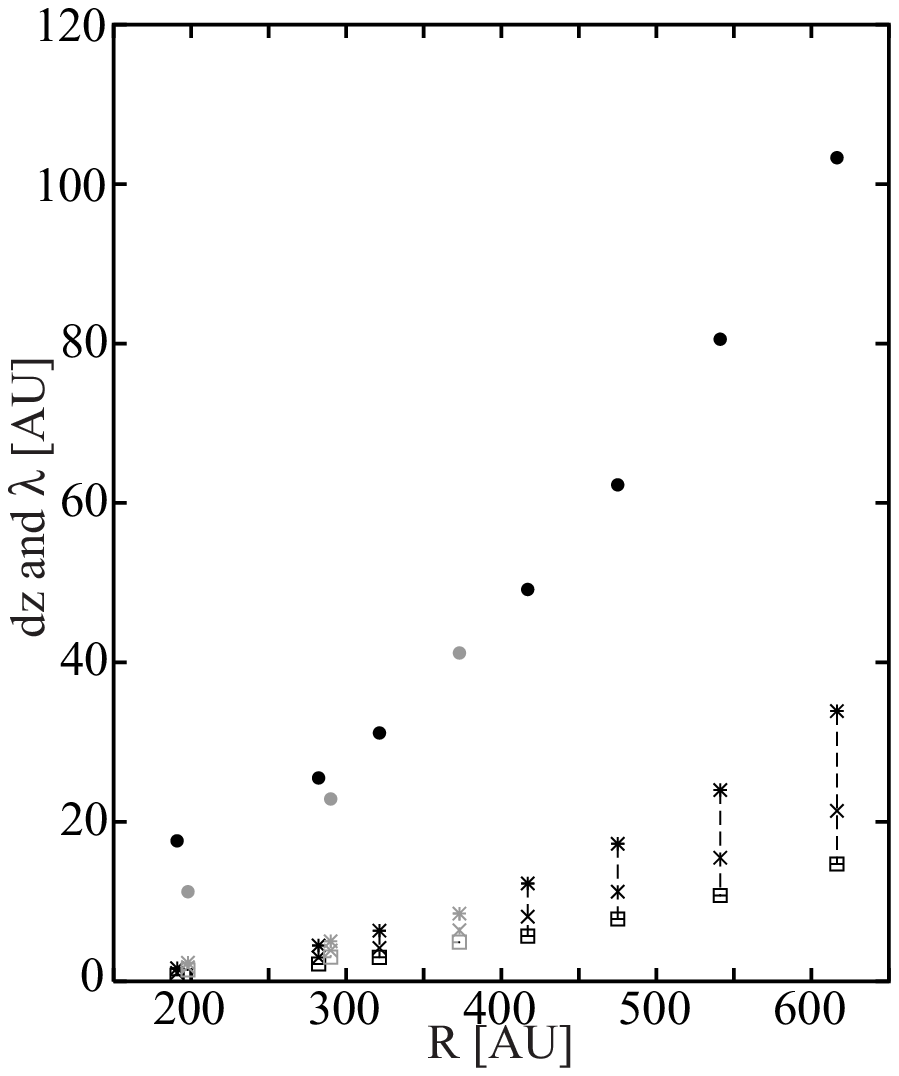}
\caption{
The black marks depict DM Tau disk model by Ceccarelli. et al. (2005),
while the gray marks are the disk model by D'Alessio et al. (1999) with the
accretion rate of $10^{-8} M_{\odot}$ yr$^{-1}$.
Asterisks, open boxes, and crosses depict $\lambda$ with the density
$n_{\rm H}$ at $T=20$ K layer, $T=13$ K layer, and with the average density
of $T=13-20$ K layer at each radius, respectively. In the DM Tau disk model,
for example,
the densities $n_{\rm H}$ at $T=13$ K and 20 K layers are $6.6\times 10^6$
cm$^{-3}$ and $1.2\times 10^6$ cm$^{-3}$ at $r=616$ AU, while they are
$4.2\times 10^8$ cm$^{-3}$ and $9.1\times 10^7$ cm$^{-3}$ at $r=191$ AU,
respectively.
Closed-circles depict the distance between the layers of $T=20$ K and 13 K
($dz_{\rm 20-13}$). \label{fig1}}
\end{figure}

\clearpage

\begin{figure}
\epsscale{1.2}
\plotone{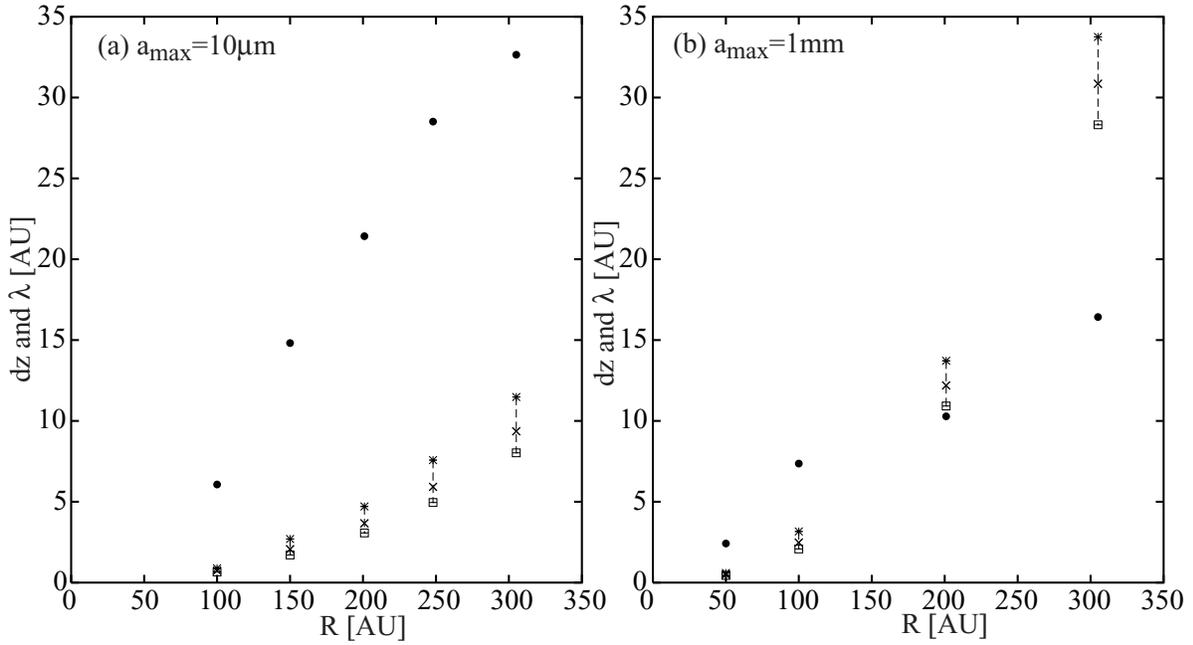}
\caption{Decay length of CO abundance $\lambda$ and distance between the layers of $T=20$ K and 13 K ($dz_{\rm 20-13}$) as in Figure 1. Disk models are by Aikawa \& Nomura (2006) with the maximum grain size of 10 $\mu$m (a) and 1 mm (b).
In the model with $a_{\rm max}=10 \mu$m (a), the densities $n_{\rm H}$ at
$T=13$ K and 20 K layers are $2.4\times 10^7$ cm$^{-3}$ and $1.2\times 10^7$
cm$^{-3}$ at $r=305$ AU, while they are $6.8\times 10^8$ cm$^{-3}$ and $3.7
\times 10^8$ cm$^{-3}$ at $r=100$ AU, respectively.
In the model with $a_{\rm max}=1$ mm (b), the density $n_{\rm H}$ at
$T=13$ K and 20 K layers are $1.9\times 10^7$ cm$^{-3}$ and $1.4\times 10^7$
cm$^{-3}$ at $r=305$ AU, while they are $5.2\times 10^9$ cm$^{-3}$ and $2.9
\times 10^9$ cm$^{-3}$ at $r=50$ AU, respectively.
\label{fig2}}
\end{figure}


\end{document}